\documentclass[cmbright]{staauth}

\usepackage{moreverb}

\usepackage[colorlinks,bookmarksopen,bookmarksnumbered,citecolor=red,urlcolor=red]{hyperref}

\usepackage{natbib}

\newcommand\BibTeX{{\rmfamily B\kern-.05em \textsc{i\kern-.025em b}\kern-.08em
T\kern-.1667em\lower.7ex\hbox{E}\kern-.125emX}}

\def\volumeyear{2016}

\runninghead{H Cho}{Unsystematic stationarity test}

\begin{document}

\theoremstyle{definition}
\newtheorem{thm}{Theorem}
\theoremstyle{definition}
\newtheorem{cor}{Corollary}
\theoremstyle{definition}
\newtheorem{lem}{Lemma}
\theoremstyle{definition}
\newtheorem{prop}{Proposition}
\theoremstyle{definition}
\newtheorem{assum}{Assumption}
\theoremstyle{remark}
\newtheorem{rem}{Remark}
\theoremstyle{definition}
\newtheorem{definition}{Definition}

\newcommand{\gam}{\gamma}
\newcommand{\del}{\delta}
\newcommand{\lam}{\lambda}
\newcommand{\sig}{\sigma}
\newcommand{\vep}{\varepsilon}

\newcommand{\pr}{\mathbb{P}}
\newcommand{\corr}{\mbox{cor}}
\newcommand{\cov}{\mbox{cov}}
\newcommand{\var}{\mbox{var}}
\newcommand{\cum}{\mbox{cum}}
\newcommand{\tr}{\mbox{tr}}
\newcommand{\TV}{\mbox{TV}}
\newcommand{\E}{\mathbb{E}}
\newcommand{\Z}{\mathbb{Z}}
\newcommand{\R}{\mathbb{R}}

\newcommand{\cA}{\mathcal{A}}
\newcommand{\cB}{\mathcal{B}}
\newcommand{\cC}{\mathcal{C}}
\newcommand{\cD}{\mathcal{D}}
\newcommand{\cJ}{\mathcal{J}}
\newcommand{\cL}{\mathcal{L}}
\newcommand{\cM}{\mathcal{M}}
\newcommand{\cN}{\mathcal{N}}
\newcommand{\cS}{\mathcal{S}}
\newcommand{\cT}{\mathcal{T}}

\newcommand{\xtt}{X_{t, T}}
\newcommand{\ijk}{I_{j, k}}
\newcommand{\zjk}{Z_{j, k}}
\newcommand{\bo}{\beta^o}
\newcommand{\js}{j^*}
\newcommand{\jts}{J^*_T}
\newcommand{\smax}{\mbox{\scriptsize{max}}}
\newcommand{\savg}{\mbox{\scriptsize{avg}}}
\newcommand{\cTm}{\cT}
\newcommand{\cTa}{\cT^{\savg}}
\newcommand{\pim}{\pi_T^{\smax}}
\newcommand{\pia}{\pi_T^{\savg}}

\newcommand{\bI}{\mathbf{I}}
\newcommand{\bR}{\mathbf{R}}
\newcommand{\bX}{\mathbf{X}}
\newcommand{\bY}{\mathbf{Y}}
\newcommand{\bpsi}{\boldsymbol{\psi}}
\newcommand{\bzero}{\mathbf{0}}
\newcommand{\bbI}{\mathbb{I}}

\def\wh{\widehat}
\def\wt{\widetilde}

\title{A test for second-order stationarity of time series based on unsystematic sub-samples}

\author{Haeran Cho \affil{a}\corrauth}

\address{%
\affilnum{a} School of Mathematics, University of Bristol, Bristol, BS8~1TW, UK\\
}

\corremail{haeran.cho@bristol.ac.uk}

\received{13 September 2016}
\accepted{3 October 2016}

\begin{abstract}
In this paper, we introduce a new method for testing the stationarity of time series, where the test statistic is obtained from measuring and maximising the difference in the second-order structure over pairs of randomly drawn intervals. The asymptotic normality of the test statistic is established for both Gaussian and a range of non-Gaussian time series, and a bootstrap procedure is proposed for estimating the variance of the main statistics. Further, we show the consistency of our test under local alternatives. Due to the flexibility inherent in the random, unsystematic sub-samples used for test statistic construction, the proposed method is able to identify the intervals of significant departure from the stationarity without any dyadic constraints, which is an advantage over other tests employing systematic designs. We demonstrate its good finite sample performance on both simulated and real data, particularly in detecting localised departure from the stationarity.
\end{abstract}


\maketitle

\section{Introduction}
\label{sec:intro}

The second-order (weak) stationarity assumption is appealing in time series analysis
as there are well-established models, estimation and forecasting tools
and accompanying asymptotic theory for statistical inference. 
In recent years, however, locally stationary time series models
such as those proposed in \cite{dahlhaus1997}, \cite{nason2000} and \cite{ombao2005}
have gained considerable attention,
since the stationarity assumption is often found unrealistic 
for long time series observed in naturally non-stationary environments.
Therefore, in order to reduce the risk of fitting a mis-specified model and generating inaccurate forecasts,
testing the validity of the stationarity assumption is an essential step in modern time series analysis.

Several statistical procedures have been proposed for stationarity testing,
since the proposal of one of the earliest tests in \cite{priestley1969}. 
\cite{rainer2000} considered a sequence of local alternatives to the stationarity 
(which converges to a stationary process at a controlled rate as the length of the time series increases),
and categorised existing stationarity tests 
into those tailored for detecting global or localised deviations, i.e.,
alternative hypotheses of the following forms
\begin{eqnarray}
\label{eq:loc:alt}
H^G_{1, T}: \, g(z) = \bar{g} + T^{-\delta} g_0(z) \mbox{ \ and \ } 
H^L_{1, T}:\, g(z) = \bar{g} + T^{-\delta} g_0(z/T^{-\gamma}), \mbox{ \ respectively},
\end{eqnarray}
where $g$ denotes a function related to the second-order structure of a time series on the rescaled interval $[0, 1]$, 
$g_0$ satisfies $\int_0^1 g_0(z)^2 dz > 0$, and $\delta, \gamma > 0$.

Stationarity tests proposed in \cite{paparoditis2009}, \cite{dette2011} and \cite{preuss2013} belong to the first category,
with $L_2$-distance or Kolmogorov-Smirnov (KS) distance-type test statistics 
that measure the quadratic deviation between the local and global spectral density estimates integrated over frequency. 
A test belonging to the second category generally involves maximising over the localised deviations from the stationarity,
which are computed from the estimates of time-evolving second-order structure that are smoothed with a certain bandwidth.
For example, the test investigated in \cite{paparoditis2010} maximises the $L_2$-distance between
the local and global spectral density estimates over a rolling window.
\cite{rainer2000} and \cite{nason2013} proposed to use Haar wavelets 
in producing systematic sub-samples over which localised departure from the stationarity is measured,
where the length of a Haar wavelet vector serves the role of a bandwidth.
\cite{wang2014} investigated the discrepancy between the local and global estimates of autocovariances at multiple lags,
adopting the Walsh functions to systematically generate the local autocovariance estimates.

In this paper, we investigate the use of {\em unsystematic} sub-samples
for capturing the localised departure from the stationarity.
It is achieved by comparing the second-order behaviour of the data over pairs of randomly drawn, disjoint intervals.
As opposed to adopting any systematic scheme such as Haar wavelets or Walsh functions,
it is expected that through the flexibility inherent in the unsystematic sub-sampling, 
the proposed test may be better suited for capturing localised non-stationarities. 
Without any dyadic constraints on the sub-samples,
our method can identify the intervals of (the most) distinctive second-order structure.
Besides, it only requires a lower bound on the size of random intervals
and thus is not encumbered with the challenging choice of window size.
As in \cite{nason2013}, we adopt the Locally Stationary Wavelet model \citep{nason2000} 
under which the asymptotic normality of the chosen statistic is established for Gaussian and a broad range of non-Gaussian processes.
It is also shown that the proposed test is consistent under local alternatives converging to the null
at a rate slower than $T^{-1/2}$, attaining the efficiency comparable to that reported in e.g., \cite{wang2014}.

The rest of the paper is organised as follows. 
In Section \ref{sec:lsw}, we describe a class of locally stationary time series 
which provides a theoretical setting for developing our test.
The proposed test statistic is introduced in Section \ref{sec:wbs},
where we establish its asymptotic normality and theoretical consistency under a sequence of local alternatives.
Also, a bootstrap procedure is introduced for handling the difficult task of estimating the unknown variance of the proposed statistics.
Section \ref{sec:fin} illustrates the comparative performance of the proposed test on simulated datasets.
Section \ref{sec:conc} concludes the paper. 
The supporting information contains all the proofs and an application to real data analysis
as well as an R package, \texttt{unsystation}, implementing the proposed test.

\section{Locally stationary wavelet model}
\label{sec:lsw}

In this section, we define the Locally Stationary Wavelet (LSW) time series model, first proposed in \cite{nason2000}. 
Here, the definition given in \cite{van2008} is presented.

\begin{definition}
\label{def:lsw} A sequence of doubly-indexed stochastic processes
$\{X_{t, T}\}_{t=0}^{T-1}$ for $T=1, 2, \ldots,$ with mean zero,
is in the class of LSW processes if there exists a representation
\begin{eqnarray}
\xtt = \sum_{j=-\infty}^{-1} \sum_{k=-\infty}^\infty w_{j, k:T} \psi_{j, t-k} \xi_{j, k}, 
\label{eq:def:lsw}
\end{eqnarray}
where $j \in \cJ = \{-1, -2, \ldots\}$ and $k \in \Z$ are the scale and location parameters, respectively, and
$\{\psi_{j, k}\}_{j, k}$ is a family of discrete, compactly supported, non-decimated wavelets.
Further, the following conditions are satisfied.
\begin{itemize}
\item[(a)] $\xi_{j, k}$ are random orthonormal increments with $\E(\xi_{j, k}) = 0$
and $\cov(\xi_{j, k}, \xi_{j', k'}) = \delta_{j, j'}\delta_{k, k'}$ for all $j, j'\in\cJ$ and $k, k'\in\Z$,
where $\delta_{a, b} = 1$ if $a=b$ and $\delta_{a, b} = 0$ otherwise.

\item[(b)] For each $j \le -1$, there exists a function $W_j(z)$ defined on $[0, 1)$ 
which possesses the following properties:
\begin{itemize}
\item[(i)] $\sum_j |W_j(z)|^2 < \infty$ uniformly in $z \in [0, 1)$;
\item[(ii)] there exists a sequence of constants $C_j$ such that for each $T$,
$\sup_{0 \le k \le T-1} \vert w_{j, k:T} - W_j(k/T)\vert \le C_j/T$;
\item[(iii)] the total variation of $W_j^2(z)$ is bounded by $L_j$;
\item[(iv)] the constants $C_j$ and $L_j$ satisfy
$\sum_{j=-\infty}^{-1} \cL_j(\cL_jL_j+C_j) < \bar{L} < \infty$,
where $\cL_j = |\bpsi_j| = (2^{-j}-1)(\cL_{-1}-1)+1$ for $\bpsi_j = (\psi_{j, 0}, \ldots, \psi_{j, \cL_j-1})^\top$ and $|\cdot|$ denotes the length of the support of a vector. 
\end{itemize}
\end{itemize}
\end{definition}

Comparing the above definition with the spectral representation of stationary processes, 
$W_j(z)$ is a scale and location-dependent transfer function, 
the wavelets $\psi_{j, t-k}$ are analogous to the Fourier exponentials, 
and the innovations $\xi_{j, k}$ correspond to the orthonormal increment process.
In what follows, we omit ``$T$'' from the subscripts for brevity when there is no confusion.

\cite{nason2000} defined the evolutionary wavelet spectrum (EWS) as $S_j(z) = W_j^2(z)$
for all $z \in [0, 1)$ and $j \in \cJ$,
which quantifies the contribution to the variance of the series at scale $j$ and time $t = \lfloor zT \rfloor$. 
For stationary processes, $S_j(z)$ remains constant for all $z$.
Through the regularity conditions imposed on the total variation of $S_j(z)$,
the LSW model in \eqref{eq:def:lsw} admits time series with both smooth and abrupt transitions in its second-order structure. 
 
The EWS is estimated using the wavelet-based local periodogram
$\ijk = |\sum_{t=0}^{T-1} X_t\psi_{j, k-t}|^2$ for $k=0, 1, \ldots, T-1$,
which is simply a sequence of the squared wavelet coefficients of the series.
Asymptotically, $\E(\ijk)$ is ``close'' to the function $\beta_j(z) = \sum_{l=-\infty}^{-1} S_l(z)A_{j, l}$
in the sense that
\begin{eqnarray}
\label{eq:per:beta}
|\E(\ijk) - \beta_j(k/T)| = O(T^{-1}) \qquad \mbox{(Proposition 3.3, \cite{nason2000})}
\end{eqnarray}
at all fixed scales $j$,
where the definition of the wavelet operator matrix $\mathbf{A} = (A_{j, l})_{j, l<0}$ can be found in \cite{nason2000}.
$\E(\ijk)$ remains constant in $k$ for stationary time series, whereas
any change in the second-order structure of $X_t$ is detectable by examining the constancy of wavelet periodograms at multiple scales.
\cite{nason2013} chose wavelet periodograms as statistics on which the Haar wavelet-based stationarity testing was performed.
For a formal argument establishing the one-to-one correspondence between wavelet periodogram sequences
and the autocovariance structure of $X_t$, see \cite{nason2000} and \cite{cho2012}. 
We note that wavelet periodogram do not require smoothing or block size selection in their computation,
and ``encode'' the localised behaviour of the second-order structure of $X_t$ in the sequences of lengths comparable to $T$ ($T - 2^{-j}+1$ to be precise).

\section{Proposed methodology}
\label{sec:wbs}

\subsection{Main statistic}
\label{sec:test:stat}

We denote a set of randomly drawn intervals within $[0, T)$ by 
$\cM = \{L_p = [s_p, e_p]: \, 0 \le s_p < e_p \le T-1, \ p=1, \ldots, M \equiv M_T\}$,
where any two end-points are randomly drawn from $\{0, \ldots, T-1\}$ with equal probability subject to a condition on 
$n_p = e_p-s_p+1$ as specified later in (A2).
The collection of the pairs of indices corresponding to any two disjoint intervals belonging to $\cM$, is denoted by
$\cD \equiv \cD(\cM) = \{(p, q); \, 1 \le p < q \le M, \, L_p \cap L_q = \emptyset\}$ with its cardinality $|\cD| = D$.
Let
\begin{eqnarray*}
\wt{\cC}_j(p, q) = \sqrt{\frac{n_pn_q}{n_p+n_q}}\left\{\frac{1}{n_p}\int_{s_p}^{e_p}\beta_j\left(\frac{y}{T}\right)dy
-\frac{1}{n_q}\int_{s_q}^{e_q}\beta_j\left(\frac{y}{T}\right)dy\right\}
\end{eqnarray*}
for any $1 \le p, q \le M$ satisfying $L_p \cap L_q = \emptyset$. 
Recalling that for stationary time series, the EWS $S_j(z)$ remain constant in $z$ for all $j$ and so $\beta_j(z)$ are,
it is evident that $\wt{\cC}_j(p, q)=0$ for all $(p, q) \in \cD$ and scale $j$ for stationary $X_t$.

As seen in (\ref{eq:per:beta}), $\ijk$ is an asymptotically 
unbiased estimator of $\beta_j(k/T)$ for each $j=-1, -2, \ldots$, 
which implies that testing the stationarity of $X_t$ 
can be achieved by evaluating the constancy of $\E(\ijk)$.
Hence, the sample counterpart of $\wt{\cC}_j(p, q)$ is defined as
\begin{eqnarray}
\cC_j(p, q) &=& 
\sqrt{\frac{n_pn_q}{n_p+n_q}}
\left(\frac{1}{n_p} \sum_{k=s_p}^{e_p} \ijk - \frac{1}{n_q} \sum_{k=s_q}^{e_q} \ijk\right).
\label{eq:test:stat:j}
\end{eqnarray}
By examining $\cC_j(p, q)$ over a sufficiently large number of pairs of randomly drawn intervals,
our hope is that 
there exists a pair of intervals $L_{p^*}$ and $L_{q^*}$ such that
any departure from constancy of $\E(\ijk)$ at some scale $j^*$,
is reflected as the large value of $|\cC_{j^*}(p^*, q^*)|$.
The correspondence between the stationarity of $X_t$ and the constancy of $\beta_j(z)$ was also utilised in \cite{nason2013}.
However, the test statistics therein are Haar wavelet coefficients of $\ijk$,
which are $\cC_j(p, q)$ with $L_p$ and $L_q$ chosen under systematic constraints,
namely that the intervals are at dyadic locations and of the same length that is a power of two.

We notice the resemblance of our approach to the Wild Binary Segmentation (WBS) proposed in \cite{piotr2014}
which computes CUSUM statistics on a collection of randomly drawn intervals for change-point analysis.
However, the WBS is developed for detecting abrupt shifts in the underlying structure of the data (e.g., mean) that is otherwise constant,
and is not suited for detecting smooth and gradual changes over time.

\subsection{Stationarity test}
\label{sec:props}

First, we establish the asymptotic properties of $\cC_j(p, q)$ under the following assumptions.
\begin{itemize}
\item[(A1)] The cumulants of $\{X_t\}_{t=0}^{T-1}$ satisfies 
$\sup_{0 \le t_1 \le T-1} \sum_{t_2, \ldots, t_k = 0}^{T-1} |\cum(X_{t_1}, X_{t_2}, \ldots, X_{t_k})| \le C_0^k(k!)^{1+\lambda}$ 
for all $k=2, 3, \ldots$ where $\lambda \ge 0$ and $C_0 > 0$.
\item[(A2)] For any $L_p \in \cM$, its length is bounded from the below,
i.e., $\min_{1 \le p \le M} n_p \ge m_T$, for some $m_T$ satisfying $(\log^2T)^{-1}m_T \to \infty$ as $T \to \infty$.
\end{itemize}
In Remark 3.1 of \cite{neumann1996}, it was shown that (A1) holds for an $\alpha$-mixing process
when (i) its mixing coefficients satisfy $\alpha(k) \le C\exp(-b|k|)$, 
and (ii) $\E|X_t|^k \le C_0^k(k!)^\lambda$ for all $k$ and some $C, C_0$ and $b > 0$.
The condition (ii) is met by many distributions such as exponential, 
gamma and inverse Gaussian besides the Gaussian distribution.

The choice of a bandwidth or window size 
is an inherent problem in most statistical inference for locally stationary processes.
\cite{rainer2000} and \cite{nason2013} impose an assumption on 
the coarseness of the Haar wavelets used for test statistic construction,
while \cite{paparoditis2010} does so on the size of the rolling window.
In \cite{wang2014}, the indices of Walsh functions are chosen such that
all the sub-intervals defined by the selected functions are of length greater than $T^{2/3}$. 
Analogously but distinctively, $m_T$ serves as a {\em lower} bound on the lengths of all the intervals in $\cM$, 
over which wavelet periodograms are ``smoothed'' (averaged) out and contrasted as in \eqref{eq:test:stat:j}.

The following lemma and proposition are comparable to Lemma 3.2, Proposition 3.1 of \cite{neumann1997} and Lemma 1, Proposition 1 of \cite{nason2013},
which are modified to accommodate the unsystematic nature of $\cC_j(p, q)$.
We note that the proof of Lemma \ref{lem:zero} is not a straightforward application of the proofs given in the other papers.

\begin{lem}
\label{lem:zero}
For $X_t$ defined in (\ref{eq:def:lsw}), let (A1)--(A2) hold.
We denote the ``wavelet'' vector associated with $\cC_j(p, q)$ by $\bpsi^{p, q} = (\psi^{p, q}_0, \ldots, \psi^{p, q}_{T-1})^\top$, where
\begin{eqnarray*}
\psi^{p, q}_k = \left\{\begin{array}{ll}
\sqrt{n_pn_q/(n_p+n_q)} n_p^{-1} & \mbox{for } s_p \le k \le e_p, \\
-\sqrt{n_pn_q/(n_p+n_q)} n_q^{-1} & \mbox{for } s_q \le k \le e_q, \\
0 & \mbox{otherwise},
\end{array}\right.
\end{eqnarray*}
such that $\cC_j(p, q) = \langle \bI, \bpsi^{p, q} \rangle$ for $\bI_j = (I_{j, 0}, \ldots, I_{j, T-1})^\top$.
Then, given $\cM$, the followings hold uniformly for $(p, q) \in \cD$ and $j \le -1$.
 \begin{itemize}
\item[(a)] $\E\{\cC_j(p, q)\} = \wt{\cC}_j(p, q) + O(m_T^{-1/2})$.
\item[(b)] $\var\{\cC_j(p, q)\} = \phi^{(1)}_j(p, q) +\phi^{(2)}_j(p, q) +\phi^{(3)}_j(p, q)$, where
$\phi^{(1)}_j(p, q)  = 2\sum_{k} (\psi^{p, q}_k)^2 \sum_u \{\sum_\tau\Psi_j(\tau-u)c(k/T, \tau)\}^2$,
$\phi^{(2)}_j(p, q) = O(\cL_j^3m_T^{-1}\log\,T)$ and $\phi^{(3)}_j(p, q) = O(1)$,
and $c(z, \tau)$ is a wavelet transform of $S_j(z)$ that denotes the local autocovariance function of $X_t$ \citep[Definition 2.9]{nason2000}.
\item[(c)] $|\cum_n(\cC_j(p, q))| \le (n!)^{2\lambda+2} K^n (\cL_jm_T^{-1/2})^{n-2}$ for all $n \ge 3$ and some fixed $K > 0$.
\end{itemize}
\end{lem}
\begin{rem}
\label{rem:sigma}
The compactness of the range of $u$ (namely, $|u| < 2\cL_j$) and the fact that $\sum_k(\psi^{p, q}_k)^2 = 1$,
lead to $\phi^{(1)}_j(p, q) \le C\cL_j$ for some $C>0$.
On the other hand, $\phi^{(1)}_j(p, q)$
$\ge 2\sum_{k} (\psi^{p, q}_k)^2 \{\sum_\tau\Psi_j(\tau)c(k/T, \tau)\}^2$
$=$ $2\sum_{k} (\psi^{p, q}_k)^2 \beta_j(k/T)^2$, 
which is bounded away from zero.
We can derive tighter bounds if, e.g., $X_t$ is time-modulated white noise defined in \cite{piotr2005} with Haar wavelets, as
$\phi^{(1)}_j(p, q)$ $= 2\sum_{k} (\psi^{p, q}_k)^2c(k/T, 0)^2 \sum_u \Psi_j^2(u)$ $\ge 2\min_zc(z, 0)^2A_{j, j} \asymp 2^{-j}$
\citep[Theorem 2.15]{nason2000}, 
indicating the dominance of $\phi^{(1)}_j(p, q)$ in $\var\{\cC_j(p, q)\}$.
However, it is difficult to generalise this result due to the presence of $\phi^{(3)}_j(p, q)$,
which arises from approximating the moments of $\cC_j(p, q)$ by those of a quadratic form of Gaussian random variables.
Nevertheless, $\var\{\cC_j(p, q)\}$ is bounded away from zero while being bounded from the above by $\cL_j$, uniformly in $j$ and $(p, q)$.
\end{rem}

From Lemma \ref{lem:zero}, it is reasonable to take into consideration wavelet periodograms at a limited number of finest scales only.
We allow $\jts$, the coarsest scale to be examined, to slowly grow with $T$ as below. 
\begin{itemize}
\item[(A3)] $\jts \to \infty$ as $T \to \infty$ subject to $\jts \le \lfloor c^*\log\log\,T\rfloor$ for $c^* \in (0, 1/3)$.
\end{itemize}
Let $a \asymp b$ indicate the existence of fixed $C, C'>0$ such that $C|b| \le |a| \le C'|b|$.
Then, the choice of $M \asymp T^2$ amounts to including every interval $[s, e]$ defined by any $1 \le s < e \le T$ in $\cM$,
which leads to examining the discrepancy between $T^4$ pairs of intervals via $\cC_j(p, q)$.
In order to avoid such a computationally intensive search, we impose the following condition on $M$. 
Later in Section \ref{sec:power}, conditions are established so that any non-stationarity can be detected without such an exhaustive search,
and the practical selection of $M$ is discussed in Appendix A of the supporting information.
\begin{itemize}
\item[(A4)] $M \asymp T^\varpi$ for some $\varpi \in (0, 2)$.
\end{itemize}

Equipped with Lemma \ref{lem:zero}, the following proposition shows the asymptotic normality of $\cC_j(p, q)$.

\begin{prop}
\label{prop:zero}
Let the assumptions in Lemma \ref{lem:zero} hold and $\Lambda_T \asymp \sqrt{\log\,T}$. Then, given $\cM$,
\begin{eqnarray*}
\pr\left\{\pm \frac{\cC_j(p, q) - \wt{\cC}_j(p, q)}{\sigma_j(p, q)} \ge x \right\} = \{1-\Phi(x)\}\{1+o(x)\}
\end{eqnarray*}
uniformly in $-\infty \le x \le \Lambda_T$ for all $(p, q) \in \cD$ and $j=-1, \ldots, -\jts$,
where $\sigma_j(p, q) = \sqrt{\var\{\cC_j(p, q)\}}$ and $\Phi(x)$ denotes the cumulative density function of the standard normal distribution.
\end{prop}

Motivated by Proposition \ref{prop:zero}, we propose the following test 
\begin{eqnarray}
\cTm := \max_{(p, q) \in \cD} \max_{-\jts \le j \le -1} \frac{|\cC_j(p, q)|}{\sigma_j(p, q)} > \Delta_T, \label{eq:test}
\end{eqnarray}
where $\Delta_T = C_\delta\sqrt{\log\,T}$ with some fixed $C_\delta > 0$.
Under the null hypothesis of stationarity,
$\wt{\cC}_j(p, q) = 0$ over any non-overlapping intervals $L_p$ and $L_q$, 
and therefore 
$\{\sigma_j(p, q)\}^{-1}\cC_j(p, q) \sim \cN(0, 1)$
for all $j = -1, \ldots, -\jts$ and $(p, q) \in \cD$.
Then, an immediate consequence of Proposition \ref{prop:zero} is 
the consistency of the proposed test (\ref{eq:test}) under the null hypothesis.

\begin{prop}
\label{prop:one}
Let $X_t$ be defined as in (\ref{eq:def:lsw}) with $S_j(z)$ constant in $z \in [0, 1)$ for all $j \le -1$, and suppose that (A1)--(A4) hold.
Then for $\Delta_T = C_\delta\sqrt{\log\,T}$ with some $C_\delta > 2\sqrt{\varpi}$,
we have $\pr(\cTm \ge \Delta_T) \to 0$ as $T \to \infty$.
\end{prop}

Using $\Delta_T$ as a test criterion requires the choice of the unknown constant $C_\delta$,
which is closely related to that of $M$ and consequently to the choices of other unknown quantities
as discussed in Section \ref{sec:power}.
Instead, noting that the test involves multiple testing over $D$ pairs of intervals and $\jts$ scales by construction, 
we adopt the Bonferroni correction and use the $(1-0.5\alpha/(D\jts)\}$-quantile of the standard normal distribution 
as a critical value at a nominal level $\alpha \in (0, 1)$.

\subsection{Consistency of the test statistic under the local alternatives}
\label{sec:power}

In this section, we study the asymptotic consistency of the proposed test under a series of local alternatives.
Motivated by the invertible relationship between the EWS and $\beta_j(z)$,
an alternative hypothesis of non-stationarity is formulated with the latter:
\begin{eqnarray}
\label{eq:loc:alt:beta}
H_{1, T}:\, \beta_j(z) = \bar{\beta}_j + \nu_T\bo_j(z)
\mbox{ with } \int_0^1 \bo_j(z)^2 dz > 0 \mbox{ for at least one scale } j=-1, -2, \ldots,
\end{eqnarray}
where $\nu_T \to 0$ as $T \to \infty$
such that the sequence of the alternatives converges to the null hypothesis at the rate $\nu_T$. 

We further assume the following conditions on $\bo_j(z)$
with the notations $A \vee B = \max(A, B)$ and $A \wedge B = \min(A, B)$.
\begin{itemize}
\item[(B1)] There exists at least one $\js \in \{-1, \ldots, -\jts\}$ for which $\int_0^1 \bo_{\js}(z)^2 dz > 0$.
\item[(B2)] For some fixed $\omega_1, \omega_2$
and $\del_T$ satisfying $(\log^2T)^{-1}\del_T \to \infty$ as $T \to \infty$, 
\begin{itemize}
\item[(a)] $\bo_{\js}(t/T)>\omega_1$ for all $t \in [a_1, b_1]$ ($0 \le a_1 < b_1 \le T-1$), and
\item[(b)] $\bo_{\js}(t/T)<\omega_2$ for all $t \in [a_2, b_2]$ ($0 \le a_2 < b_2 \le T-1$),
\end{itemize}
where $\omega_1 > \omega_2$, $(b_1-a_1+1) \wedge (b_2-a_2+1) \ge \del_T$ and $[a_1, b_1] \cap [a_2, b_2] = \emptyset$.
\end{itemize}
Under (A3), (B1) follows trivially from (B2) as noted in Appendix B.1 of \cite{cho2015}.
(B2) assumes the existence of two disjoint intervals over which $\bo_{\js}(z)$ is greater (smaller) 
than a fixed constant $\omega_1$ ($\omega_2$),
without imposing any further restrictions on its shape.
For example, (B2) is satisfied when $\bo_{\js}(z)$ is piecewise constant 
with at least a single jump that is bounded away from zero and located at a sufficient distance from the both ends.
In such a context, $\del_T$ is related to the minimum spacing of change-points.
Besides, (B2) is met, for example, when $\bo_{\js}(z)$ is a sinusoidal signal with a fixed amplitude and 
its frequency (number of oscillations) increasing with $T$ at the rate $T\del_T^{-1}$ or slower. 
On the other hand, when the departure from the constancy of $\bo_{\js}(z)$ 
takes the form of a spike (e.g., a mixture of Dirac delta functions), 
then $\del_T \asymp 1$ and hence (B2) is violated. 
We further make the following assumption.
\begin{itemize}
\item[(B3)] $(\log\,T)^{-1}\nu_T \del_T^{1/2} \to \infty$ as $T \to \infty$. 
\end{itemize}

The rate at which the local alternatives are allowed to converge to the null hypothesis is dependent 
on the length of the intervals of non-stationarities through $\del_T$. 
Due to this dependence, although the local alternatives in (\ref{eq:loc:alt:beta}) are presented in the form of $H^G_{1, T}$ in (\ref{eq:loc:alt}),
they are comparable to $H^L_{1, T}$ implicitly.
If $\del_T \asymp T$, (B3) boils down to $\nu_T \gg T^{-1/2}\log\,T$,
which is competitive with the rates provided by \cite{preuss2013} and \cite{wang2014} up to a factor logarithmic in $T$.
On the other hand, when $\nu_T = 1$ (fixed alternative), $\del_T$ is permitted to be logarithmic in $T$.
Under the above conditions, our test is asymptotically consistent under the sequence of local alternatives in (\ref{eq:loc:alt:beta}).
\begin{thm}
\label{thm:one}
Let (A1)--(A4) and (B1)--(B3) hold, and $m_T  = c\delta_T$ for some $c \in (0, 1)$. Then,
\begin{eqnarray}
\pr(\cTm > \Delta_T) &\ge& 1 - 
\frac{2\sqrt{\log\,T}}{\sqrt{2\pi}\nu_T\delta_T^{1/2}(\omega_1-\omega_2)}\exp\left(-\frac{\nu_T^2\delta_T(\omega_1-\omega_2)^2}{8\log\,T}\right) 
- 2\left\{1-\frac{c(1-c)\delta_T^2}{T^{2}}\right\}^M \nonumber
\\
&\ge&
1 - C T^{-\varepsilon} - 2\left\{1-\frac{c(1-c)\delta_T^2}{T^{2}}\right\}^M \to 1 \quad \mbox{as} \quad T \to \infty \label{eq:thm:one}
\end{eqnarray}
for some fixed $C, \varepsilon > 0$.
\end{thm}

\subsection{Estimation of $\sigma_j(p, q)$}
\label{sec:sigma}

Estimation of the scaling term $\sigma_j(p, q)$ is an enduring problem
as its estimator $\wh\sigma_j(p, q)$ plays an important role 
in ensuring that, our test based on
$\wh{\cTm} := \max_{(p, q) \in \cD} \max_{-\jts \le j \le -1} \{\wh\sigma_j(p, q)\}^{-1}|\cC_j(p, q)|$
and the critical value selected via Bonferroni correction,
performs well even for small sample size.

Motivated by Lemma \ref{lem:zero} (b) and the definition of $c(z, \tau)$, 
we may approximate $\phi^{(1)}_j(p, q)$ via estimating the EWS.
However, as discussed in Remark \ref{rem:sigma}, the contribution of $\phi^{(3)}_j(p, q)$ to $\var\{\cC_j(p, q)\}$ may be non-negligible.
Moreover, while there are a few procedures proposed for estimating the EWS, 
some rely on the assumption of piecewise stationarity \citep{piotr2006}
or provide only pointwise estimates of the wavelet spectrum \citep{van2008}.
The R package \texttt{wavethresh} \citep{wavethresh} offers a readily applicable estimation procedure,
but it occasionally returns EWS estimates with negative ordinates and may be sensitive to the choice of wavelets. 
Instead, we propose to employ the AR sieve bootstrap (see \cite{kreiss2011} and references therein) 
for estimating $\sigma_j(p, q)$ as below.
\begin{description}
\item[AR sieve bootstrap for estimating $\sigma_j(p, q)$.]
\item[Step 1:] An AR model of order $p$ is fitted to the observed data $X_0, \ldots, X_{T-1}$, 
where $p \equiv p_T$ increases with the sample size $T$ while $p \ll T$.
The Yule-Walker estimators of the AR parameters are denoted by $\wh{\alpha}_1, \ldots, \wh{\alpha}_p$,
and the residuals are $\wt{Z}_t = X_t - \sum_{i=1}^p \wh{\alpha}_i X_{t-i}$.
Finally, the centred residuals are obtained as $\wh{Z}_t = \wt{Z}_t - \bar{Z}$ with $\bar{Z} = (T-p)^{-1}\sum_{t=p}^{T-1}\wt{Z}_t$.

\item[Step 2:] For $b = 1, \ldots, B$, repeat the following steps.
\begin{description}
\item[Step 2.1:] Draw $Z^b_t, \, t=0, \ldots, T$ independently from $\wh{F}_T$, the empirical distribution of $\wh{Z}_t$,
and let $X_t^b = \sum_{i=1}^p \wh{\alpha}_i X_{t-i}^b + Z_t^b$.
\item[Step 2.2:] Compute 
$\cC^b_j(p, q) = \sqrt{\frac{n_pn_q}{n_p+n_q}} (\frac{1}{n_p} \sum_{k=s_p}^{e_p} \ijk^b - \frac{1}{n_q} \sum_{k=s_q}^{e_q} \ijk^b)$
for all $j=-1, \ldots, -\jts$ and $(p, q) \in \cD$, where $\ijk^b = |\sum_{t=0}^{T-1} X^b_t\psi_{j, k-t}|^2$.
\end{description}

\item[Step 3:] $\sigma_j(p, q)$ is estimated by 
$\wh{\sigma}_j(p, q) = B^{-1} \sum_{b=1}^B \{\cC^b_j(p, q) - \bar{\cC}_j(p, q)\}^2$,
where $\bar{\cC}_j(p, q) = B^{-1} \sum_{b=1}^B \cC^b_j(p, q)$.
\end{description}

\section{Finite sample performance}
\label{sec:fin}

In the simulations below, we consider several stationary and non-stationary time series models. 
For each model, we generated $R = 100$ replications with varying sample size ($T = 256, 512, 1024$).
Some practical guidance on the choice of $\cM$, wavelets for computing $\ijk$, 
the coarsest wavelet scale $\jts$ and $m_T$ is provided in Appendix A (available as supporting information for this paper).
We apply the proposed test with these quantities chosen according to the guidance and 
the bootstrap procedure in Section \ref{sec:sigma} with the bootstrap sample size $B = 200$.
For comparison, we have included the tests proposed in
\cite{nason2013} \citep[R package \texttt{locits}]{locits}, 
\cite{preuss2013}, \cite{preuss2014} and \cite{wang2014} (denoted by HWTOS, PVD, PP and JWW, respectively),
each of which was applied with the tuning parameters chosen as per the suggestions made in the respective papers.

\subsection{Size of the test}
\label{sec:sim:null}

We investigate the empirical Type I errors of our test on the data simulated from the following ARMA models 
first studied in \cite{nason2013},
with innovations drawn from the standard normal, Gamma$(9, 1)$ and $t_5$ distributions.

\begin{itemize}
\item[(S1)] White noise model.
\item[(S2)] AR(1) process with the AR parameter $-0.9$. 
\item[(S3)] AR(1) process with the AR parameter $0.9$. 
\item[(S4)] MA(1) process with the MA parameter $-0.8$. 
\item[(S5)] MA(1) process with the MA parameter $0.8$. 
\item[(S6)] ARMA(1, 2) process with the AR parameter $-0.4$ and MA parameters $-0.8, 0.4$. 
\item[(S7)] AR(2) process with the AR parameters $1.385929, -0.9604$. 
\end{itemize}
The results are summarised in Tables \ref{table:sim:null:normal}--\ref{table:sim:null:t},
which report the proportion of rejections at given significance level $\alpha = 0.1$ and $0.05$ under each model.

With the standard normal innovations, 
our test performs well in keeping the empirical Type I errors below the nominal level or slightly above for all (S1)--(S7).
In (S2), the large, negative AR parameter leads to 
Haar wavelet periodograms with high autocorrelations at scale $j = -1$,
which in turn leads to Haar wavelet coefficients computed on the periodograms being spuriously large, 
and this is reflected as the large Type I error returned by the HWTOS.
As for (S7), the process is stationary but close to a unit root process
with a spectral peak around $\pi/4$ (see \cite{nason2013} for more detail)
and as a result, the PVD, PP, JWW and HWTOS struggle to keep the size below the nominal levels. 
Moreover, the HWTOS tends not to scale well with increasing sample size, 
which is also observed with different innovation distributions (see Tables \ref{table:sim:null:gamma}--\ref{table:sim:null:t}).

In Table \ref{table:sim:null:gamma}, the proposed test performs similarly well 
with the innovations drawn from the Gamma distribution.
The performance of HWTOS is similar to that observed from Table \ref{table:sim:null:normal}.
On the other hand, the PVD and PP rely on the Gaussianity of the time series, 
which plays an essential role in their proposed bootstrap procedures for test criterion selection,
and hence they tend to perform poorly with non-Gaussian innovations.
Most tends to perform poorly with increasing sample size when the innovations are drawn from the $t_5$-distribution.
$\wh\cTm$ suffers from the heavy tail of the innovation distribution that violates (A1),  and so do the HWTOS, PVD and PP.
The JWW shows marginally better size performance for some models, such as (S1) and (S5)--(S6), but 
performs worse than $\wh{\cTm}$ for (S3) and (S7).

\subsection{Power of the test}
\label{sec:sim:power}

We investigate the empirical power of the proposed test
on a number of non-stationary models:
(N1)--(N4) in \cite{nason2013}, (N5)--(N8) in \cite{preuss2013} and (N9)--(N11) in \cite{wang2014}.
Additionally, we include (N12) where a piecewise stationary AR($1$) process has its AR coefficient switch between $0.5$ and $-0.5$ 
at a higher frequency with increasing $T$, so that the changes may be ``masked'' within the systematic sub-samples 
taken by the HWTOS or JWW.
In all models, $\{Z_t\}_{t=1}^T$ are generated as independent and identically distributed standard normal random variables.
\begin{itemize}
\item[(N1)] $X_t = \alpha_t X_{t-1} + Z_t$,
where $\alpha_t$ evolves linearly from $0.9$ to $-0.9$ over $t=1, \ldots, T$.
\item[(N2)] An LSW process based on Haar wavelets with spectrum 
$S_{-1}(z) = 0.25-0.5(z-0.5)^2$ and $S_j(z) = 0$ for all $j<-1$. 
\item[(N3)] An LSW process where
$S_{-1}(z)$ is as in (N2), $S_{-2}(z) = S_{-1}(z+0.5)$ using periodic boundaries (for the construction of the spectrum only)
and $S_j(z) = 0$ for all $j<-2$.
\item[(N4)] An LSW process with spectrum
$S_{-1} = \exp\{-64(z-0.5)^2\}$, $S_{-3}(z) = S_{-1}(z-0.25)$, $S_{-4}(z) = S_{-1}(z+0.25)$
and $S_j(z) = 0$ for $j=-2$ and $j<-4$. 
\item[(N5)] $X_t = (1+t/T)Z_t$.
\item[(N6)] $X_t =  -0.9\sqrt{t/T}X_{t-1} + Z_t$.
\item[(N7)] $X_t = 0.8\cos\{1.5-\cos(4\pi t/T)\}Z_{t-1} + Z_t$.
\item[(N8)] $X_t = 0.8\cos\{1.5-\cos(4\pi t/T)\}Z_{t-6} + Z_t$.
\item[(N9)] $X_t = 0.6\sin(4\pi t/T)X_{t-1} + Z_t$.
\item[(N10)] $X_t = 0.5X_{t-1} + Z_t$ for $\{1 \le t \le T/4\} \cup \{3T/4 < t \le T\}$, and 
$X_t = -0.5X_{t-1} + Z_t$ for $T/4  < t \le 3T/4$.
\item[(N11)] $X_t = -0.5X_{t-1} + Z_t$ for $\{1 \le t \le T/2\} \cup \{T/2+T/64 < t \le T\}$, and
$X_t = 4Z_t$ for $T/2  < t \le T/2+T/64$.
\item[(N12)] $X_t =  -0.5X_{t-1} + Z_t$ for $(2k)\sqrt{T} \le t \le (2k+1)\sqrt{T}$, and
$X_t = 0.5X_{t-1} + Z_t$ for $(2k+1)\sqrt{T} \le t \le (2k+2)\sqrt{T}$, where $k=0, 1, \ldots$.
\end{itemize}

Empirical power at $\alpha = 0.1$ and $0.05$ attained by the different tests is reported in Table \ref{table:sim:power}.
The test criterion chosen via Bonferroni correction for $\wh{\cTm}$ behaves rather conservatively, 
as evidenced by the lower power observed for smaller samples.
However, the power performance of our test improves progressively as $T$ increases beyond $T \ge 512$.

Overall, the JWW attains the largest power for many models,
but our test is shown to be much better suited for detecting more localised departure from the stationarity, such as those exhibited in (N11)--(N12).
Also, compared to the HWTOS, the unsystematic sub-sampling adopted by our test proves to be better in terms of its detection power.
Note that the spuriously large empirical size of the HWTOS when $T = 1024$ (see Table \ref{table:sim:null:normal}), 
needs to be taken into consideration when interpreting its large power.
In general, different methods perform particularly well for different models;
for example, the PVD outperforms our test for (N5) and (N6),
while the opposite is the case for (N7)--(N12).

\begin{figure}[htbp]
\centering
\includegraphics[scale=0.45]{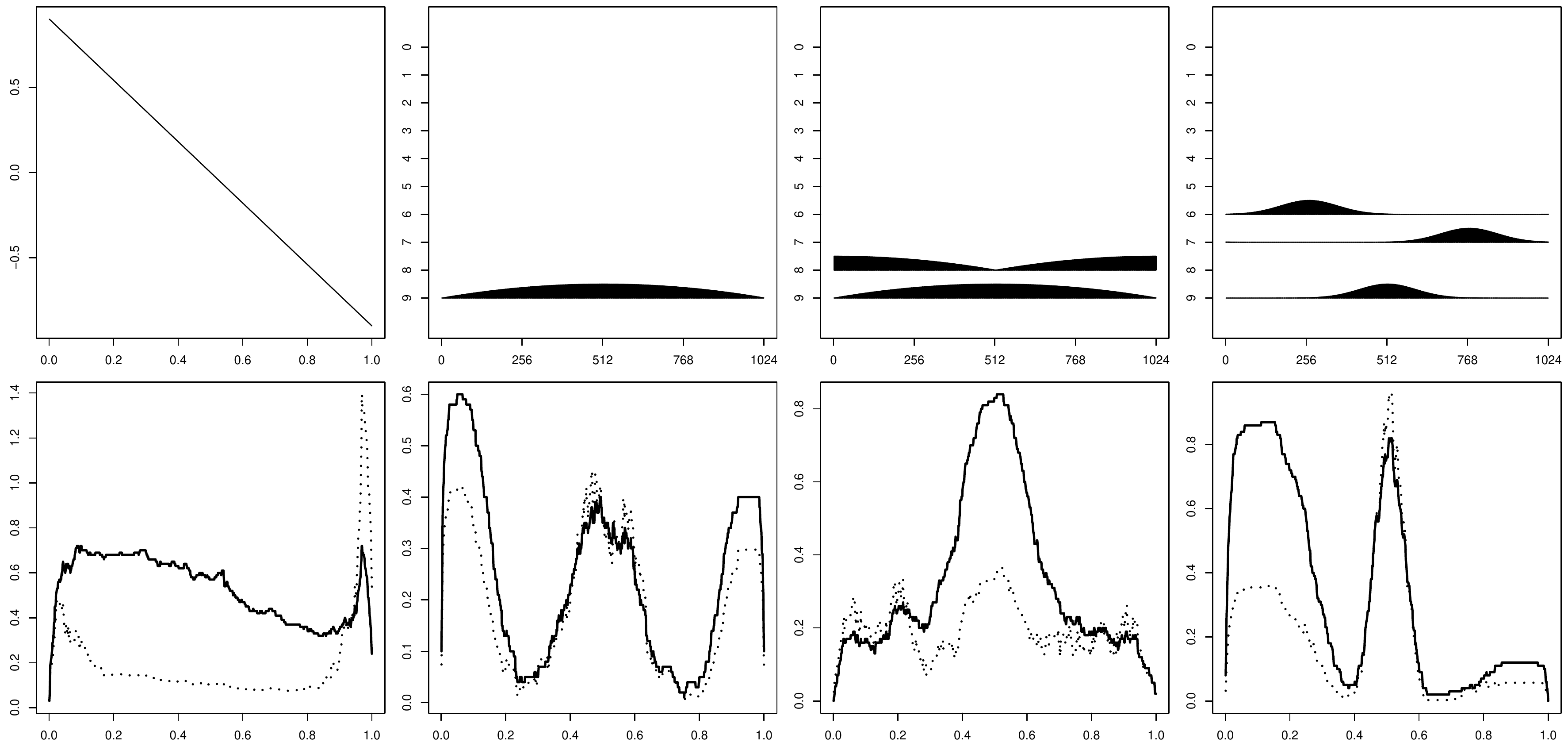}
\caption{Top: time-varying AR parameter in (N1) and scale-by-scale $S_j(z)$ in (N2)--(N4) ($j=-1, \ldots, -J_T$ from bottom to top); 
bottom: weights applied to  $L_{\wh{p}}$ and $L_{\wh{q}}$ according to the weighting schemes (i) (solid) and (ii) (dotted), 
respectively, averaged pointwise over 100 replications; $T=1024$ is used.}
\label{fig:nason:weight}
\end{figure}

\begin{figure}[htbp]
\centering
\includegraphics[scale=0.45]{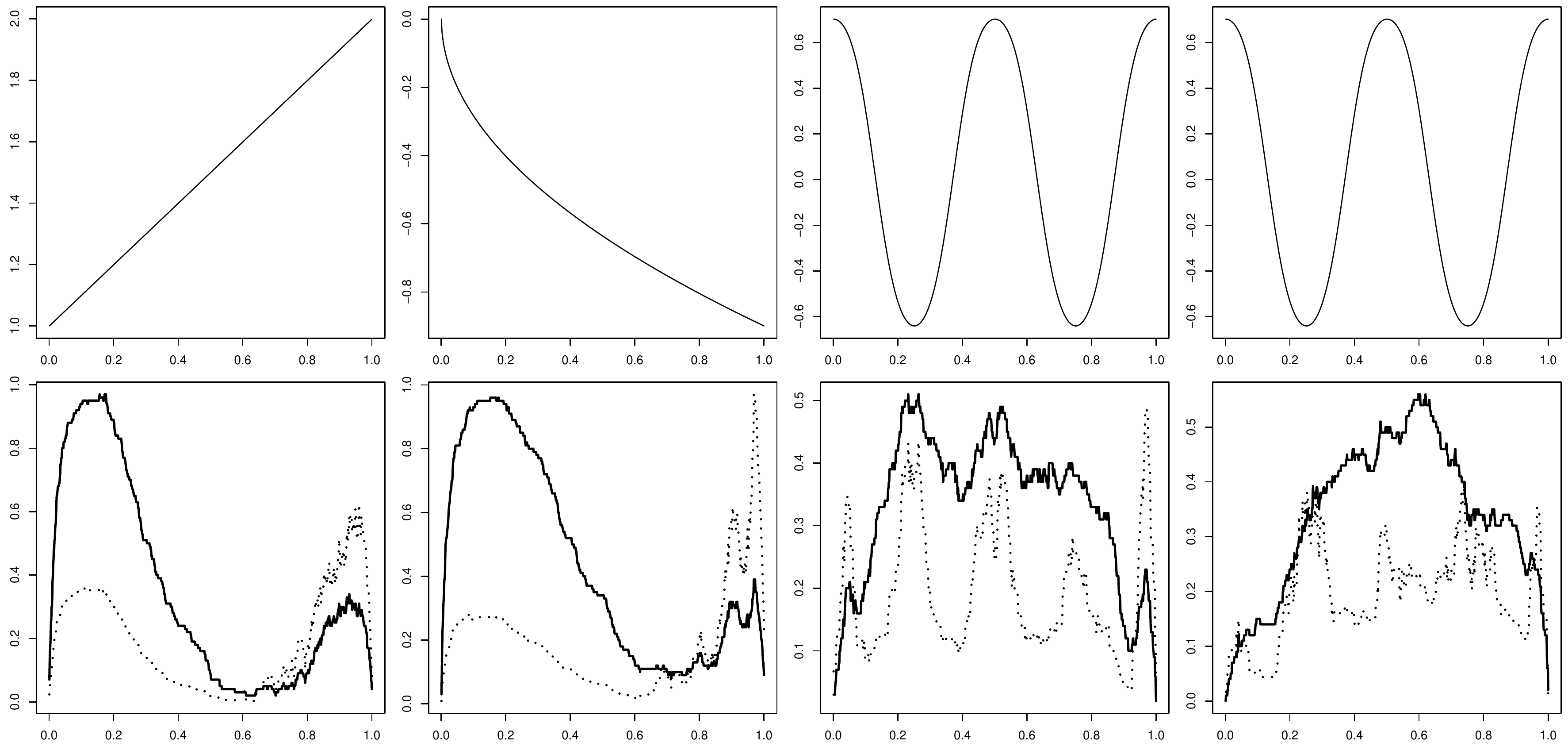}
\caption{Top: time-varying variance in (N5), time-varying AR parameter in (N6) and time-varying MA parameter in (N7)--(N8);
bottom: as in Figure \ref{fig:nason:weight}; $T=1024$ is used.}
\label{fig:dette:weight}
\end{figure}

\begin{figure}[htbp]
\centering
\includegraphics[scale=0.45]{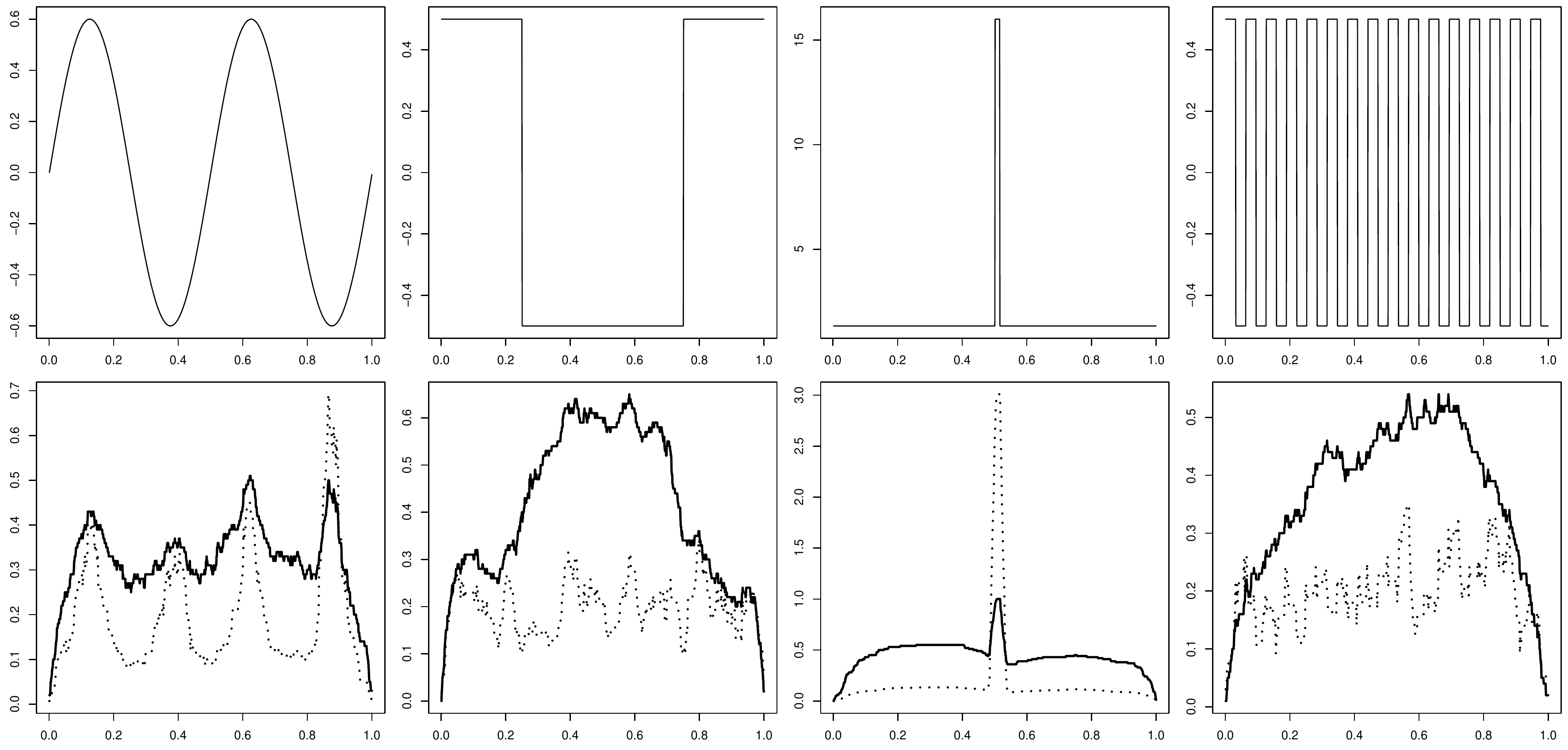}
\caption{Top: time-varying AR parameter in (N9)--(N10), time-varying variance in (N11) and time-varying AR parameter in (N12);
bottom: as in Figure \ref{fig:nason:weight}; $T=1024$ is used.}
\label{fig:wang:weight}
\end{figure}

Our localised approach allows us to identify the intervals of the most distinctive second-order structure in the observed time series,
through the maximisation taken over $(p, q) \in \cD$,
without any dyadic restrictions on both their locations and lengths.
We confirm this by plotting the weights given to the intervals $L_{\wh{p}}$ and $L_{\wh{q}}$, 
where $(\wh{p}, \wh{q}) = \arg\max_{(p, q) \in \cD} \max_{-\jts \le j \le -1} \{\wh\sigma_j(p, q)\}^{-1}|\cC_j(p, q)|$,
averaged pointwise over $R = 100$ realisations for each model
in Figures \ref{fig:nason:weight}--\ref{fig:wang:weight},
according to two weighting schemes:
(i) the equal weight of $1/R$ is assigned to every point in both intervals, and 
(ii) the weights reciprocal to the lengths ($1/n_{\wh{p}}$ and $1/n_{\wh{q}}$) are assigned to $L_{\wh{p}}$ and $L_{\wh{q}}$.
Employing the weighting scheme (ii), we can learn about the length of the chosen intervals as well as their locations.

The locations of peaks and troughs in the quantities representing the time-varying second-order structure,
coincide with those of the sharp peaks formed in the weights given to $L_{\wh{p}}$ and $L_{\wh{q}}$,
which is particularly noticeable in (N7) and (N9).
Also, the sharp spike in the weights plotted for (N11) indicates that the brief interval over which 
$X_t$ switches from being a stationary AR($1$) process to being a white noise series and back,
is clearly captured by our approach.
In (N4)--(N5), the time-varying quantities exhibit increasing or decreasing trends,
and our method correctly identifies the intervals of the most dissimilar autocovariance structure
by preferring the intervals from the two extreme ends.

\subsection{Real data analysis}
\label{sec:real}

We illustrate the application of the proposed test to 
the weekly egg price series observed at a German agricultural market
between April 1967 and May 1990 in Appendix B, which is available as supporting information for this paper.

\section{Conclusions}
\label{sec:conc}

In this paper, we proposed a novel stationarity test for time series data,
which is based on pairs of randomly and unsystematically drawn sub-samples. 
Our test enables us to locate the intervals of the most distinctive second-order behaviour without any dyadic constraint,  
and achieves asymptotic consistency under a sequence of local alternatives that converge to the null hypothesis 
at a rate comparable to the efficiency of the state-of-the-art.
Accompanying the main statistic, we introduced a bootstrap scheme for estimating its variance,
which has been shown empirically to perform well under both the null and alternative hypotheses.

As multivariate, even high-dimensional time series are frequently observed,
a stationarity test applicable to such data is increasingly in demand.
We note that, working under the multivariate version of the LSW model proposed in \cite{cho2015},
it may be possible to extend our test to accommodate the high dimensionality of the data,
by combining the randomised, unsystematic sub-sampling with 
the thresholding principle proposed therein or the Double CUSUM operator of \cite{cho2016a}.
We leave this for future investigation.

\subsection*{Supporting information}

Additional information for this article is available where a practical guide for choice of parameters, 
an application of the proposed test to real data analysis and detailed proofs of theoretical results can be found.
It also includes an implementation of the proposed test as an R package \texttt{unsystation},
which has been submitted and will be made available on the Comprehensive R Archive Network archive.

\newpage

\bibliographystyle{wb_stat}
\bibliography{fbib}

\begin{thebibliography}{24}
\newcommand{\enquote}[1]{`#1'}
\providecommand{\natexlab}[1]{#1}
\expandafter\ifx\csname urlstyle\endcsname\relax
  \providecommand{\doi}[1]{doi:\discretionary{}{}{}#1}\else
  \providecommand{\doi}{doi:\discretionary{}{}{}\begingroup
  \urlstyle{rm}\Url}\fi

\bibitem[{Cho(2016)}]{cho2016a}
Cho, H (2016), \enquote{Change-point detection in panel data via double cusum
  statistic,} \emph{Electronic Journal of Statistics}, \textbf{10}, pp.
  2000--2038.

\bibitem[{Cho \& Fryzlewicz(2012)}]{cho2012}
Cho, H \& Fryzlewicz, P (2012), \enquote{Multiscale and multilevel technique
  for consistent segmentation of nonstationary time series,} \emph{Statistica
  Sinica}, \textbf{22}, pp. 207--229.

\bibitem[{Cho \& Fryzlewicz(2015)}]{cho2015}
Cho, H \& Fryzlewicz, P (2015), \enquote{Multiple change-point detection for
  high-dimensional time series via sparsified binary segmentation,}
  \emph{Journal of the Royal Statistical Society, Series B}, \textbf{77}, pp.
  475--507.

\bibitem[{Dahlhaus(1997)}]{dahlhaus1997}
Dahlhaus, R (1997), \enquote{Fitting time series models to nonstationary
  processes,} \emph{Annals of Statistics}, \textbf{25}, pp. 1--37.

\bibitem[{Dette et~al.(2011)Dette, Preu{\ss} \& Vetter}]{dette2011}
Dette, H, Preu{\ss}, P \& Vetter, M (2011), \enquote{A measure of stationarity
  in locally stationary processes with applications to testing,} \emph{Journal
  of the American Statistical Association}, \textbf{106}, pp. 1113--1124.

\bibitem[{Fryzlewicz(2005)}]{piotr2005}
Fryzlewicz, P (2005), \enquote{Modelling and forecasting financial log-returns
  as locally stationary wavelet processes,} \emph{Journal of Applied
  Statistics}, \textbf{32}, pp. 503--528.

\bibitem[{Fryzlewicz(2014)}]{piotr2014}
Fryzlewicz, P (2014), \enquote{Wild binary segmentation for multiple
  change-point detection,} \emph{Annals of Statistics}, \textbf{42}, pp.
  2243--2281.

\bibitem[{Fryzlewicz \& Nason(2006)}]{piotr2006}
Fryzlewicz, P \& Nason, GP (2006), \enquote{Haar--{F}isz estimation of
  evolutionary wavelet spectra,} \emph{Journal of the Royal Statistical
  Society: Series B}, \textbf{68}, pp. 611--634.

\bibitem[{Jin et~al.(2014)Jin, Wang \& Wang}]{wang2014}
Jin, L, Wang, S \& Wang, H (2014), \enquote{A new non-parametric stationarity
  test of time series in the time domain,} \emph{Journal of the Royal
  Statistical Society: Series B}, \textbf{77}, pp. 893--922.

\bibitem[{Kreiss et~al.(2011)Kreiss, Paparoditis \& Politis}]{kreiss2011}
Kreiss, JP, Paparoditis, E \& Politis, DN (2011), \enquote{On the range of
  validity of the autoregressive sieve bootstrap,} \emph{Annals of Statistics},
  \textbf{39}, pp. 2103--2130.

\bibitem[{Nason(2013{\natexlab{a}})}]{locits}
Nason, G (2013{\natexlab{a}}), \emph{locits: Tests of stationarity and
  localized autocovariance}, {R} package version 1.4.

\bibitem[{Nason(2013{\natexlab{b}})}]{wavethresh}
Nason, G (2013{\natexlab{b}}), \emph{wavethresh: Wavelets statistics and
  transforms}, {R} package version 4.6.6.

\bibitem[{Nason(2013{\natexlab{c}})}]{nason2013}
Nason, GP (2013{\natexlab{c}}), \enquote{A test for second-order stationarity
  and approximate confidence intervals for localized autocovariances for
  locally stationary time series,} \emph{Journal of the Royal Statistical
  Society: Series B}, \textbf{75}, pp. 879--904.

\bibitem[{Nason et~al.(2000)Nason, {v}on Sachs \& Kroisandt}]{nason2000}
Nason, GP, {v}on Sachs, R \& Kroisandt, G (2000), \enquote{Wavelet processes
  and adaptive estimation of the evolutionary wavelet spectrum,} \emph{Journal
  of the Royal Statistical Society: Series B}, \textbf{62}, pp. 271--292.

\bibitem[{Neumann(1996)}]{neumann1996}
Neumann, MH (1996), \enquote{Spectral density estimation via nonlinear wavelet
  methods for stationary non-gaussian time series,} \emph{Journal of Time
  Series Analysis}, \textbf{17}, pp. 601--633.

\bibitem[{Neumann \& {v}on Sachs(1997)}]{neumann1997}
Neumann, MH \& {v}on Sachs, R (1997), \enquote{Wavelet thresholding in
  anisotropic function classes and application to adaptive estimation of
  evolutionary spectra,} \emph{Annals of Statistics}, \textbf{25}, pp. 38--76.

\bibitem[{Ombao et~al.(2005)Ombao, {v}on Sachs \& Guo}]{ombao2005}
Ombao, H, {v}on Sachs, R \& Guo, W (2005), \enquote{{SLEX} analysis of
  multivariate nonstationary time series,} \emph{Journal of the American
  Statistical Association}, \textbf{100}, pp. 519--531.

\bibitem[{Paparoditis(2009)}]{paparoditis2009}
Paparoditis, E (2009), \enquote{Testing temporal constancy of the spectral
  structure of a time series,} \emph{Bernoulli}, \textbf{15}, pp. 1190--1221.

\bibitem[{Paparoditis(2010)}]{paparoditis2010}
Paparoditis, E (2010), \enquote{Validating stationarity assumptions in time
  series analysis by rolling local periodograms,} \emph{Journal of the American
  Statistical Association}, \textbf{105}, pp. 839--851.

\bibitem[{Preu{\ss} et~al.(2013)Preu{\ss}, Vetter \& Dette}]{preuss2013}
Preu{\ss}, P, Vetter, M \& Dette, H (2013), \enquote{A test for stationarity
  based on empirical processes,} \emph{Bernoulli}, \textbf{19}, pp. 2715--2749.

\bibitem[{Priestley \& {Subba Rao}(1969)}]{priestley1969}
Priestley, M \& {Subba Rao}, T (1969), \enquote{A test for non-stationarity of
  time-series,} \emph{Journal of the Royal Statistical Society: Series B},
  \textbf{31}, pp. 140--149.

\bibitem[{Puchstein \& Preu{\ss}(2016)}]{preuss2014}
Puchstein, R \& Preu{\ss}, P (2016), \enquote{Testing for stationarity in
  multivariate locally stationary processes,} \emph{Journal of Time Series
  Analysis}, \textbf{37}, pp. 3--29.

\bibitem[{Van~Bellegem \& {v}on Sachs(2008)}]{van2008}
Van~Bellegem, S \& {v}on Sachs, R (2008), \enquote{Locally adaptive estimation
  of evolutionary wavelet spectra,} \emph{Annals of Statistics}, \textbf{36},
  pp. 1879--1924.

\bibitem[{{v}on Sachs \& Neumann(2000)}]{rainer2000}
{v}on Sachs, R \& Neumann, MH (2000), \enquote{A wavelet-based test for
  stationarity,} \emph{Journal of Time Series Analysis}, \textbf{21}, pp.
  597--613.

\end{thebibliography}

\newpage

\begin{table}[htb]
\caption{Empirical size at the significance level $\alpha = 0.1, 0.05$ with $\cN(0, 1)$ innovations.}
\label{table:sim:null:normal}
\centering
{\footnotesize
\begin{tabular}{c c | c c | c c | c c | c c | c c}
\toprule											
&	 &
\multicolumn{2}{c}{$\wh\cTm$} &		\multicolumn{2}{c}{HWTOS} 	&	\multicolumn{2}{c}{PVD} &		\multicolumn{2}{c}{PP} &		\multicolumn{2}{c}{JWW} 	\\	
$T$ &	$\alpha$ &	0.1 &	0.05 &	0.1 &	0.05 &	0.1 &	0.05 &	0.1 &	0.05 &	0.1 &	0.05	\\	\midrule
256 &	(S1) &	0.04 &	0.01 &	0.08 &	0.03 &	0.13 &	0.07 &	0.18 &	0.1 &	0.16 &	0.09	\\	
&	(S2) &	0.08 &	0.08 &	0.29 &	0.2 &	0.11 &	0.03 &	0.12 &	0.05 &	0.06 &	0.04	\\	
&	(S3) &	0.02 &	0.02 &	0.06 &	0.02 &	0.07 &	0.02 &	0.14 &	0.09 &	0.05 &	0.03	\\	
&	(S4) &	0.02 &	0.02 &	0 &	0 &	0.09 &	0.03 &	0.08 &	0.01 &	0.14 &	0.06	\\	
&	(S5) &	0.02 &	0.02 &	0.13 &	0.05 &	0.12 &	0.06 &	0.08 &	0.03 &	0.13 &	0.04	\\	
&	(S6) &	0.05 &	0.04 &	0.05 &	0.02 &	0.1 &	0.07 &	0.11 &	0.04 &	0.07 &	0	\\	
&	(S7) &	0.03 &	0.02 &	0.12 &	0.12 &	0.26 &	0.14 &	0.22 &	0.11 &	0.18 &	0.1	\\	\midrule
													
512 &	(S1) &	0.05 &	0.05 &	0.08 &	0.02 &	0.13 &	0.06 &	0.12 &	0.06 &	0.09 &	0.07	\\	
&	(S2) &	0.03 &	0.03 &	0.26 &	0.16 &	0.08 &	0.02 &	0.09 &	0.01 &	0.07 &	0.05	\\	
&	(S3) &	0.04 &	0.04 &	0.1 &	0.05 &	0.05 &	0.03 &	0.08 &	0.02 &	0.09 &	0.04	\\	
&	(S4) &	0.02 &	0.02 &	0.01 &	0 &	0.11 &	0.04 &	0.15 &	0.06 &	0.08 &	0.04	\\	
&	(S5) &	0.04 &	0.04 &	0.07 &	0.03 &	0.12 &	0.06 &	0.13 &	0.07 &	0.14 &	0.09	\\	
&	(S6) &	0.05 &	0.05 &	0.01 &	0.01 &	0.09 &	0.04 &	0.14 &	0.06 &	0.07 &	0.04	\\	
&	(S7) &	0.05 &	0.05 &	0.27 &	0.23 &	0.15 &	0.05 &	0.15 &	0.07 &	0.17 &	0.11	\\	\midrule
													
1024 &	(S1) &	0.06 &	0.06 &	0.39 &	0.29 &	0.08 &	0.01 &	0.15 &	0.07 &	0.1 &	0.06	\\	
&	(S2) &	0.05 &	0.04 &	0.35 &	0.25 &	0.11 &	0.04 &	0.07 &	0.01 &	0.07 &	0.02	\\	
&	(S3) &	0.05 &	0.03 &	0.22 &	0.11 &	0.08 &	0.05 &	0.1 &	0.05 &	0.08 &	0.04	\\	
&	(S4) &	0.04 &	0.03 &	0.07 &	0.06 &	0.07 &	0.05 &	0.07 &	0.06 &	0.1 &	0.03	\\	
&	(S5) &	0.05 &	0.02 &	0.28 &	0.19 &	0.04 &	0.03 &	0.11 &	0.03 &	0.12 &	0.06	\\	
&	(S6) &	0.03 &	0.02 &	0.05 &	0.04 &	0.1 &	0.07 &	0.16 &	0.06 &	0.07 &	0.05	\\	
&	(S7) &	0.05 &	0.02 &	0.3 &	0.24 &	0.07 &	0.06 &	0.06 &	0.02 &	0.19 &	0.14	\\	\bottomrule
\end{tabular}}
\end{table}

\begin{table}[htb]
\caption{Empirical size at the significance level $\alpha = 0.1, 0.05$ with Gamma$(9, 1)$ innovations.}
\label{table:sim:null:gamma}
\centering
{\footnotesize
\begin{tabular}{c c | c c | c c | c c | c c | c c}
\toprule											
&	 &	
\multicolumn{2}{c}{$\wh\cTm$} &		\multicolumn{2}{c}{HWTOS} 	&	\multicolumn{2}{c}{PVD} &		\multicolumn{2}{c}{PP} &		\multicolumn{2}{c}{JWW} 	\\	
$T$ &	$\alpha$ &	0.1 &	0.05 &	0.1 &	0.05 &	0.1 &	0.05 &	0.1 &	0.05 &	0.1 &	0.05	\\	\midrule
256 &	(S1) &	0.05 &	0.04 &	0.14 &	0.11 &	0.07 &	0.06 &	0.87 &	0.79 &	0.02 &	0.02	\\	
&	(S2) &	0.07 &	0.06 &	0.3 &	0.24 &	0.17 &	0.09 &	0.05 &	0 &	0.06 &	0.04	\\	
&	(S3) &	0.06 &	0.04 &	0.07 &	0.02 &	0.02 &	0.01 &	0.57 &	0.55 &	0.02 &	0.01	\\	
&	(S4) &	0.02 &	0.02 &	0 &	0 &	0.14 &	0.1 &	0.08 &	0.06 &	0.09 &	0.04	\\	
&	(S5) &	0.05 &	0.05 &	0.12 &	0.04 &	0.01 &	0 &	0.38 &	0.35 &	0.02 &	0.02	\\	
&	(S6) &	0.04 &	0.04 &	0.01 &	0 &	0.21 &	0.09 &	0.08 &	0.04 &	0.03 &	0.02	\\	
&	(S7) &	0.02 &	0.02 &	0.08 &	0.06 &	0.24 &	0.1 &	0.16 &	0.1 &	0.08 &	0.04	\\	\midrule
													
512 &	(S1) &	0.03 &	0.02 &	0.08 &	0.02 &	0.09 &	0.02 &	0.81 &	0.75 &	0.06 &	0.03	\\	
&	(S2) &	0.08 &	0.08 &	0.18 &	0.14 &	0.16 &	0.11 &	0.1 &	0.03 &	0.12 &	0.05	\\	
&	(S3) &	0.05 &	0.03 &	0.12 &	0.04 &	0.03 &	0.01 &	0.71 &	0.7 &	0.04 &	0.03	\\	
&	(S4) &	0.02 &	0.02 &	0.05 &	0.02 &	0.13 &	0.06 &	0.13 &	0.07 &	0.09 &	0.05	\\	
&	(S5) &	0.04 &	0.04 &	0.14 &	0.04 &	0.09 &	0.02 &	0.51 &	0.45 &	0.03 &	0.02	\\	
&	(S6) &	0.06 &	0.05 &	0 &	0 &	0.15 &	0.09 &	0.09 &	0.07 &	0.08 &	0.05	\\	
&	(S7) &	0.06 &	0.06 &	0.21 &	0.18 &	0.14 &	0.06 &	0.12 &	0.04 &	0.13 &	0.09	\\	\midrule
													
1024 &	(S1) &	0.09 &	0.07 &	0.32 &	0.2 &	0.15 &	0.09 &	0.85 &	0.8 &	0.03 &	0.02	\\	
&	(S2) &	0.09 &	0.07 &	0.32 &	0.23 &	0.13 &	0.06 &	0.1 &	0.03 &	0.16 &	0.09	\\	
&	(S3) &	0.08 &	0.08 &	0.43 &	0.3 &	0.01 &	0 &	0.87 &	0.84 &	0 &	0	\\	
&	(S4) &	0.06 &	0.05 &	0.09 &	0.08 &	0.16 &	0.09 &	0.18 &	0.09 &	0.07 &	0.03	\\	
&	(S5) &	0.09 &	0.07 &	0.35 &	0.26 &	0.08 &	0.02 &	0.57 &	0.53 &	0.03 &	0.02	\\	
&	(S6) &	0.08 &	0.08 &	0.08 &	0.02 &	0.19 &	0.12 &	0.15 &	0.08 &	0.11 &	0.03	\\	
&	(S7) &	0.06 &	0.06 &	0.4 &	0.32 &	0.17 &	0.09 &	0.15 &	0.03 &	0.2 &	0.12	\\	\bottomrule
\end{tabular}}
\end{table}

\begin{table}[htb]
\caption{Empirical size at the significance level $\alpha = 0.1, 0.05$ with $t_5$ innovations.}
\label{table:sim:null:t}
\centering
{\footnotesize
\begin{tabular}{c c | c c | c c | c c | c c | c c}
\toprule													
&	 &	
\multicolumn{2}{c}{$\wh\cTm$} &		\multicolumn{2}{c}{HWTOS} 	&	\multicolumn{2}{c}{PVD} &		\multicolumn{2}{c}{PP} &		\multicolumn{2}{c}{JWW} 	\\	
$T$ &	$\alpha$ &	0.1 &	0.05 &	0.1 &	0.05 &	0.1 &	0.05 &	0.1 &	0.05 &	0.1 &	0.05	\\	\midrule
256 &	(S1) &	0.05 &	0.04 &	0.19 &	0.1 &	0.3 &	0.16 &	0.27 &	0.22 &	0.08 &	0.04	\\	
&	(S2) &	0.05 &	0.04 &	0.24 &	0.19 &	0.1 &	0.03 &	0.09 &	0.03 &	0.14 &	0.08	\\	
&	(S3) &	0.06 &	0.04 &	0.13 &	0.1 &	0.22 &	0.14 &	0.18 &	0.09 &	0.08 &	0.04	\\	
&	(S4) &	0.06 &	0.06 &	0.04 &	0.02 &	0.27 &	0.15 &	0.24 &	0.16 &	0.08 &	0.06	\\	
&	(S5) &	0.15 &	0.14 &	0.18 &	0.1 &	0.33 &	0.24 &	0.28 &	0.16 &	0.06 &	0.02	\\	
&	(S6) &	0.15 &	0.14 &	0.09 &	0.08 &	0.24 &	0.14 &	0.17 &	0.14 &	0.1 &	0.06	\\	
&	(S7) &	0.05 &	0.03 &	0.12 &	0.08 &	0.2 &	0.09 &	0.15 &	0.07 &	0.1 &	0.03	\\	\midrule
													
512 &	(S1) &	0.14 &	0.12 &	0.21 &	0.14 &	0.45 &	0.26 &	0.25 &	0.19 &	0.07 &	0.03	\\	
&	(S2) &	0.14 &	0.12 &	0.32 &	0.27 &	0.13 &	0.05 &	0.13 &	0.09 &	0.13 &	0.08	\\	
&	(S3) &	0.12 &	0.11 &	0.15 &	0.09 &	0.17 &	0.08 &	0.16 &	0.11 &	0.2 &	0.1	\\	
&	(S4) &	0.08 &	0.08 &	0.1 &	0.07 &	0.36 &	0.3 &	0.3 &	0.2 &	0.1 &	0.07	\\	
&	(S5) &	0.14 &	0.12 &	0.21 &	0.1 &	0.29 &	0.15 &	0.21 &	0.14 &	0.08 &	0.05	\\	
&	(S6) &	0.16 &	0.14 &	0.05 &	0.03 &	0.21 &	0.17 &	0.18 &	0.11 &	0.1 &	0.07	\\	
&	(S7) &	0.06 &	0.05 &	0.29 &	0.21 &	0.15 &	0.09 &	0.13 &	0.04 &	0.14 &	0.09	\\	\midrule
													
1024 &	(S1) &	0.21 &	0.21 &	0.52 &	0.45 &	0.55 &	0.43 &	0.33 &	0.26 &	0.1 &	0.04	\\	
&	(S2) &	0.27 &	0.26 &	0.46 &	0.36 &	0.15 &	0.08 &	0.2 &	0.09 &	0.23 &	0.14	\\	
&	(S3) &	0.18 &	0.15 &	0.45 &	0.34 &	0.24 &	0.18 &	0.15 &	0.06 &	0.35 &	0.29	\\	
&	(S4) &	0.17 &	0.15 &	0.34 &	0.27 &	0.37 &	0.3 &	0.25 &	0.16 &	0.1 &	0.04	\\	
&	(S5) &	0.25 &	0.22 &	0.61 &	0.54 &	0.44 &	0.28 &	0.35 &	0.25 &	0.11 &	0.03	\\	
&	(S6) &	0.26 &	0.23 &	0.19 &	0.17 &	0.25 &	0.14 &	0.16 &	0.08 &	0.14 &	0.09	\\	
&	(S7) &	0.14 &	0.12 &	0.47 &	0.41 &	0.15 &	0.05 &	0.14 &	0.08 &	0.28 &	0.17	\\	\bottomrule
\end{tabular}}
\end{table}

\begin{table}[htbp]
\caption{Empirical power at the significance level $\alpha = 0.1, 0.05$.}
\label{table:sim:power}
\centering{\footnotesize
\begin{tabular}{c c | c c | c c | c c | c c | c c }
\toprule													
&	&		\multicolumn{2}{c}{$\wh\cTm$} &	 \multicolumn{2}{c}{HWTOS} &		\multicolumn{2}{c}{PVD} &		\multicolumn{2}{c}{PP} &		\multicolumn{2}{c}{JWW}	\\	
$T$ &	$\alpha$ &	0.1 &	0.05 &	0.1 &	0.05 &	0.1 &	0.05 &	0.1 &	0.05 &	0.1 &	0.05	\\	\midrule
256 &	(N1) &	0.89 &	0.89 &	0.88 &	0.71 &	0.9 &	0.78 &	0.6 &	0.37 &	1 &	1	\\	
&	(N2) &	0.24 &	0.21 &	0.05 &	0.03 &	0.41 &	0.26 &	0.08 &	0.03 &	0.72 &	0.53	\\	
&	(N3) &	0.1 &	0.1 &	0.04 &	0.04 &	0.13 &	0.08 &	0.09 &	0.02 &	0.46 &	0.33	\\	
&	(N4) &	0.98 &	0.98 &	0.67 &	0.52 &	0.79 &	0.7 &	0.22 &	0.11 &	1 &	1	\\	
&	(N5) &	0.35 &	0.3 &	0.14 &	0.07 &	0.97 &	0.95 &	0.95 &	0.95 &	0.91 &	0.82	\\	
&	(N6) &	0.43 &	0.41 &	0.34 &	0.25 &	0.81 &	0.64 &	0.81 &	0.71 &	0.76 &	0.62	\\	
&	(N7) &	0.49 &	0.47 &	0.35 &	0.22 &	0.21 &	0.09 &	0.17 &	0.1 &	1 &	1	\\	
&	(N8) &	0.11 &	0.09 &	0.09 &	0.05 &	0.09 &	0.05 &	0.1 &	0.05 &	0.22 &	0.16	\\	
&	(N9) &	0.65 &	0.63 &	0.51 &	0.29 &	0.2 &	0.12 &	0.14 &	0.09 &	1 &	1	\\	
&	(N10) &	0.56 &	0.53 &	0.59 &	0.32 &	0.23 &	0.18 &	0.13 &	0.07 &	1 &	1	\\	
&	(N11) &	0.61 &	0.57 &	0.48 &	0.37 &	0.3 &	0.17 &	0.4 &	0.26 &	0.35 &	0.16	\\	
&	(N12) &	0.22 &	0.18 &	0.09 &	0.01 &	0.23 &	0.17 &	0.15 &	0.13 &	0.18 &	0.13	\\	\midrule
																									
512 &	(N1) &	1 &	1 &	1 &	1 &	1 &	0.99 &	0.91 &	0.86 &	1 &	1	\\	
&	(N2) &	0.57 &	0.53 &	0.23 &	0.1 &	0.76 &	0.59 &	0.1 &	0.05 &	0.97 &	0.96	\\	
&	(N3) &	0.24 &	0.17 &	0.06 &	0.03 &	0.21 &	0.11 &	0.03 &	0.01 &	0.88 &	0.81	\\	
&	(N4) &	1 &	1 &	0.99 &	0.98 &	1 &	1 &	0.39 &	0.2 &	1 &	1	\\	
&	(N5) &	0.69 &	0.63 &	0.42 &	0.3 &	1 &	1 &	1 &	1 &	0.99 &	0.97	\\	
&	(N6) &	0.76 &	0.74 &	0.8 &	0.7 &	0.95 &	0.93 &	0.97 &	0.95 &	1 &	0.97	\\	
&	(N7) &	0.83 &	0.8 &	0.71 &	0.56 &	0.19 &	0.12 &	0.16 &	0.09 &	1 &	1	\\	
&	(N8) &	0.29 &	0.27 &	0.27 &	0.14 &	0.24 &	0.11 &	0.19 &	0.08 &	0.22 &	0.14	\\	
&	(N9) &	0.89 &	0.88 &	0.79 &	0.62 &	0.38 &	0.21 &	0.13 &	0.08 &	1 &	1	\\	
&	(N10) &	0.9 &	0.85 &	0.92 &	0.81 &	0.53 &	0.36 &	0.18 &	0.15 &	1 &	1	\\	
&	(N11) &	0.96 &	0.95 &	0.88 &	0.76 &	0.55 &	0.42 &	0.53 &	0.45 &	0.6 &	0.54	\\	
&	(N12) &	0.42 &	0.36 &	0.03 &	0.03 &	0.24 &	0.17 &	0.17 &	0.1 &	0.22 &	0.13	\\	\midrule
																									
1024 &	(N1) &	1 &	1 &	1 &	1 &	1 &	1 &	0.99 &	0.97 &	1 &	1	\\	
&	(N2) &	0.96 &	0.92 &	0.89 &	0.74 &	0.99 &	0.99 &	0.36 &	0.02 &	1 &	1	\\	
&	(N3) &	0.26 &	0.25 &	0.09 &	0.05 &	0.38 &	0.26 &	0.13 &	0.09 &	1 &	1	\\	
&	(N4) &	1 &	1 &	1 &	1 &	1 &	1 &	0.73 &	0.56 &	1 &	1	\\	
&	(N5) &	0.97 &	0.95 &	0.97 &	0.92 &	1 &	1 &	1 &	1 &	1 &	1	\\	
&	(N6) &	0.95 &	0.94 &	1 &	1 &	1 &	1 &	1 &	1 &	1 &	1	\\	
&	(N7) &	1 &	0.98 &	0.99 &	0.98 &	0.38 &	0.26 &	0.17 &	0.12 &	1 &	1	\\	
&	(N8) &	0.46 &	0.37 &	0.72 &	0.59 &	0.27 &	0.2 &	0.2 &	0.08 &	0.52 &	0.38	\\	
&	(N9) &	1 &	1 &	1 &	1 &	0.77 &	0.56 &	0.27 &	0.16 &	1 &	1	\\	
&	(N10) &	1 &	1 &	1 &	1 &	0.78 &	0.52 &	0.16 &	0.06 &	1 &	1	\\	
&	(N11) &	0.99 &	0.99 &	0.99 &	0.99 &	0.74 &	0.62 &	0.77 &	0.64 &	0.89 &	0.8	\\	
&	(N12) &	0.48 &	0.46 &	0.73 &	0.53 &	0.24 &	0.16 &	0.18 &	0.12 &	0.29 &	0.13	\\	\bottomrule
\end{tabular}}
\end{table}

\end{document}